\documentclass{ws-procs975x65}

\newcommand{\prd}{Phys. Rev. D}
\newcommand\unit[2]{\mbox{#1\,#2}}
\newcommand\mathunit[2]{#1\,#2}

\begin{document}

\title{SEARCHES FOR INSPIRAL GRAVITATIONAL WAVES ASSOCIATED WITH SHORT GAMMA-RAY BURSTS IN LIGO'S FIFTH AND VIRGO'S FIRST SCIENCE RUN}

\author{ALEXANDER DIETZ for the LIGO Scientific Collaboration and Virgo Collaboration}

\address{LAPP, 9 Chemin de Bellevue,\\
BP 110, 74941 Annecy le Vieux CEDEX, France\\
E-mail: alexander.dietz@lapp.in2p3.fr}


\bodymatter

\vspace*{5mm}
\begin{abstract}
This brief report describes the search for gravitational-wave inspiral signals from short gamma-ray bursts.
Since these events are probably created by the merger of two compact objects, a targeted search with a lower threshold can be made. 
The data around 22 short gamma-ray bursts have been analyzed.
\end{abstract}

\vspace*{5mm}
Gamma Ray Bursts (GRB) are intense flashes of $\gamma$-rays that occur in two classes; long duration bursts (duration $\gtrsim$ \unit{2}{s}) which are generally associated with hypernova explosions in star-forming galaxies
\cite{Campana:2006} and short duration bursts (duration $\lesssim$ \unit{2}{s}), thought to originate primarily from the coalescence of a neutron star (NS) with another
compact object\cite{NakarReview:2007}, like a neutron star or a black hole (BH).
Such coalescences produce gravitational waves (GW) which are searched for, and because the GRBs give the time and sky location the search can be performed with reduced threshold, increasing the sensitivity compared to untriggered searches. 

This is a brief report on a search for gravitational wave signals associated with short GRBs (SGRB) during LIGO's fifth science run S5 and Virgo's first science run, VSR1. 
S5 took place from 4 November 2005 to 30 September 2007 and VSR1 from 18 May 2007 to 30 September 2007.
S5 is comprised of three detectors (H1 and H2 at Hanford, WA, and L1 at Livingston, LA) and VSR1 is comprised of the Virgo detector near Pisa, Italy. 
There are 212 GRBs in total observed during this time, of which 33 are SGRBs\footnote[1]{Two of them have a duration greater than 2 seconds, but spectral features hint to a merger nature of these events.}. Because we require coincident data from at least two detectors (which are the ones with the best directional sensitivity to the known sky location of the GRB), only 22 SGRBs have been selected for the analysis.  
There are 9 GRBs in H1--H2 coincident time, 11 in H1--L1, 1 in H2--L1, and 1 GRB in H1--L1--V1 coincident time. 
The latter GRB (070923) is a special case because it uses data from three spatial separated detectors, which all have about the same directional sensitivity to GRB~070923.
Worth special mention is also GRB~070201, which has been already analyzed in a high-priority search because the progenitor was possibly located in M31, a galaxy only $\sim$780~kpc away. 
No gravitational-wave signal was found and a coalescence scenario could be ruled out with $>$99\% confidence at that distance \cite{GRB070201}. 
This GRB is being reanalyzed in the current work because of the use of a lower threshold and improved analysis methods.

The coalescence model suggests that the time delay between the arrival of the gravitational wave and the subsequent electromagnetic burst, referred to as trigger time, is of order one second or less. 
If the initial spike of GRBs is created by so-called internal shocks (produced by the impact of a faster outmoving shells on a slower moving shells in a GRB), the time-delay is on the millisecond scale \cite{NakarReview:2007,meszaros:2006} or at least below 1 second \cite{Finn:2004}. 
With a more semi-analytical description of the  final stage of a NS-–BH merger it has been calculated that the majority of matter plunges on the BH within 1 second \cite{davies:2005} and numerical simulations on the mass transfer suggest a timescales of milliseconds \cite{Rosswog:2005} or some seconds at maximum \cite{Faber:2006tx}. 
These considerations prompted the use of a \unit{6}{s} long time window (\textit{on-source}) around the trigger time (from -5 to +1), in which we expect the GW signal to be.

The data around the on-source segment is chosen to estimate the background and to perform software injections. 
To prevent a possible loud on-source signal biasing the background estimation, we discard \unit{48}{s} on each side of the on-source segment, and split the remaining data in up to 324 segments, each \unit{6}{s} long. 
The number of used off-source segments varies because some might be vetoed due to bad data quality \cite{abbott-2009}.
The data are analyzed using the standard compact binary search pipeline described in detail in [\citen{Collaboration:2009tt}], using a matched-filter technique to filter the data against post-Newtonian approximants of the expected GW signal. 
The mass-range of binaries searched for spans the range $[\mathunit{1}{M_\odot}, \mathunit{40}{M_\odot})$ in total mass. 
After having applied signal-based vetoes and performing the coincidence criterion, we end up
with a list of \textit{candidate events}; see [\citen{S5GRB}] for details.

To obtain the search efficiency and to calculate the likelihood of a candidate event, simulated waveform signals (\textit{software injections}) are added to the off-source data and processed with the pipeline described above. 
The efficiency is determined by the fractional number of recovered injections, made over a wide range of parameters (i.e. masses, spins etc.).
The likelihood ratio of a candidate is then the efficiency divided by the false-alarm probability, which is obtained by the relative number of off-source trials yielding a statistical significance equal or greater than the on-source measurement.

The analysis yielded no evidence for a gravitational-wave signal in coincidence
with any GRB in our sample. With the null observations and the large number of
injection trials we are able to constrain the distance to each GRB assuming it was
caused by a compact binary coalescence consisting of a neutron star and a neutron
star or a black hole. Detailed results on the exclusion distances are given in~[\citen{S5GRB}].

In addition to the individual detection searches, we assess the presence of a gravitational wave signal too weak to stand out above background separately, but which is significant when the entire sample of analyzed GRBs are taken together.
This is being done with the non-parametric Wilcoxon-Mann-Whitney U test \cite{MannWhitney}, to answer the question of whether both samples (significances of the on-source and off-source segments) are drawn from the same population. 
This test calculates a U statistic by the sums of the ranks of both samples, and compares it to the expected value if both samples were drawn from the same distribution. 
Since this statistic is approximately Gaussian distributed, a one-sided probability $p$ can be calculated that either shows the two samples are indeed drawn from the same distribution ($p\simeq50\%$) or contain a population of weak signals in the sample ($p\lesssim5\%$). 
A large value for $p$ would hint of a problem with the analysis, since it implies more signiﬁcant candidates in the off-source than in the on-source.
The full details of this analysis and discussion of the results can be found in~[\citen{S5GRB}].

\section*{Acknowledgements}

The authors gratefully acknowledge the support of the United States
National Science Foundation for the construction and operation of the
LIGO Laboratory, the Science and Technology Facilities Council of the
United Kingdom, the Max-Planck-Society, and the State of
Niedersachsen/Germany for support of the construction and operation of
the GEO600 detector, and the Italian Istituto Nazionale di Fisica
Nucleare and the French Centre National de la Recherche Scientifique
for the construction and operation of the Virgo detector. The authors
also gratefully acknowledge the support of the research by these
agencies and by the Australian Research Council, the Council of
Scientific and Industrial Research of India, the Istituto Nazionale di
Fisica Nucleare of Italy, the Spanish Ministerio de Educaci\'on y
Ciencia, the Conselleria d'Economia Hisenda i Innovaci\'o of the
Govern de les Illes Balears, the Foundation for Fundamental Research
on Matter supported by the Netherlands Organisation for Scientific
Research, the Royal Society, the Scottish Funding Council, the
Scottish Universities Physics Alliance, The National Aeronautics and
Space Administration, the Carnegie Trust, the Leverhulme Trust, the
David and Lucile Packard Foundation, the Research Corporation, and
the Alfred P. Sloan Foundation.
This document has been assigned LIGO document number P0900307.

\bibliographystyle{ws-procs975x65}

\end{document}